\documentstyle[11pt]{article}
\title{\bf{Rigidity and Normal Modes in Random \\ Matrix Spectra}}
      
\author{A. Andersen, A. D. Jackson, and H. J. Pedersen \\
         {\it The Niels Bohr Institute, Blegdamsvej 17, DK-2100 Copenhagen \O,
               Denmark} }

\begin{document}
\maketitle

\begin{abstract}
We consider the Gaussian ensembles of random matrices and describe the 
normal modes of the eigenvalue spectrum, i.e., the correlated fluctuations of 
eigenvalues about their most probable values. The associated normal mode 
spectrum is linear, and for large matrices, the normal modes are found to be 
Chebyshev polynomials of the second kind.  We contrast this 
with the behaviour of a sequence of uncorrelated levels, which has a quadratic 
normal mode spectrum. The difference in the rigidity of random matrix
spectra and sequences of uncorrelated levels can be attributed to this 
difference in the normal mode spectra.  We illustrate this by calculating the 
number variance in the two cases.
\end{abstract}

\section{Introduction}

Random matrices have been used to describe level correlations in many 
different areas of physics from nuclear levels to acoustic resonances to 
QCD. See, e.g., the review \cite{review}. The most important feature 
distinguishing the eigenvalues of a random matrix from a sequence of 
uncorrelated levels is the existence of strong repulsion between neighbouring 
eigenvalues.  This repulsion is conveniently pictured by the Coulomb gas 
model adopted by Dyson in 1962 \cite{dyson}.  Dyson exploited the equivalence 
between the Gaussian ensembles 
of random matrices and a classical problem in electrostatics in which 
identical uniformly charged parallel wires are placed on a line in a 
harmonic confining external field. The eigenvalues of the random matrix 
experience a repulsion which is analogous to the Coulomb 
repulsion between the line charges.  This level repulsion leads to 
the rigidity of random matrix spectra, which is most easily illustrated 
by the fact that the variance of the number of eigenvalues in an 
(unfolded) interval of length $L$ grows logarithmically with $L$.  This is in 
contrast to the linear behaviour of the number variance for a sequence of 
uncorrelated levels.

All spectral properties of random matrices can be calculated from the 
joint pro\-bability distribution of the eigenvalues.  Given the Coulomb 
analogy, it is natural to consider the independent ``normal modes'' 
of the joint probability density, which describe the correlated motion of 
eigenvalues about their most probable values.  Evidently, these normal 
modes have a simple interpretation in the Coulomb analogy where they 
describe the independent oscillations of charges on a lattice about their  
equilibrium positions.  An eigenvalue can be associated with each normal 
mode, and these eigenvalues form the normal mode spectrum.  
In the present case, soft (hard) modes correspond to large (small) 
amplitude fluctuations in the random matrix spectrum.  These collective 
degrees of freedom prove to be useful in determining the long-range 
spectral fluctuation measures of random matrices including the number 
variance mentioned above.   
    
The purpose of this paper is to calculate the normal modes of the Gaussian 
random matrix ensembles and to use them for a determination of the 
number variance.  For comparison, we will also determine the 
corresponding normal modes and the number variance for a sequence of 
uncorrelated levels.  We emphasise that the results of this exercise are 
neither new nor exact.  Our intention is rather to offer a new way to regard 
the fluctuations in random matrix spectra and to emphasise the value 
of thinking about the correlated motion of eigenvalues.   

\section{Normal Modes for the Gaussian Ensembles}

The Gaussian ensembles of $N \! \times \!N$ matrices have a joint probability 
distribution of the eigenvalues $x_{1}, x_{2}, \ldots , x_{N}$ which is
\cite{mehta}   
\begin{equation}
  P_{N \beta}(x_{1},x_{2}, \ldots , x_{N}) = 
    C_{N\beta} \prod_{1 \leq i < j \leq N} |x_{i}-x_{j}|^{\beta} 
    \exp \left(- \frac{\beta}{2} N \sum_{i=1}^{N} x_{i}^{2} \right) \ ,
\end{equation}
where $\beta = 1, 2, 4$ corresponds to the Gaussian orthogonal, unitary, and 
symplectic ensembles, respectively.  This distribution leads to an 
average level density $\rho(x)$ which is independent of $\beta$. In the 
large-$N$ limit: 
\begin{equation}
\label{leveldens.eqn}
  \rho(x) = \frac{N}{\pi} \sqrt{2 - x^{2}} \ .  
\end{equation}

The maxima of $P_{N \beta}$ correspond to the most probable locations of the
eigenvalues.  It is sufficient to consider the single maximum with 
$x_i < x_{i+1}$, and we denote the maximum value of $P_{N \beta}$ by 
$P_{N \beta}^{0}$.  One immediately obtains the following set of equations 
which determine the equilibrium positions of the eigenvalues
\begin{equation}
\label{equlibrium.eqn}
  \sum_{j \neq i} \frac{1}{x_{i}-x_{j}} - N x_{i} = 0 \ .
\end{equation}

In the vicinity of the maximum, we approximate the logarithm of $P_{N \beta}$ 
by 
\begin{equation}
\label{approximation.eqn}
  \ln P_{N \beta} = \ln P_{N \beta}^{0} + \frac{1}{2} \beta 
  \sum_{i,j} \delta x_{i} C_{ij} \delta x_{j} \ .
\end{equation}
The matrix $C$ is defined as
\begin{equation}
C_{ij} = \frac{1}{\beta} \ 
         \frac{\partial^{2}}{\partial x_{i} \partial x_{j}} 
\ln P_{N \beta} 
\end{equation}
evaluated at the maximum.
The elements of $C$ are 
\begin{eqnarray}
\label{cmatrix1.eqn}
  C_{ii} &=& - \sum_{j \neq i} \frac{1}{(x_{i} - x_{j})^{2}} - N \\ 
\label{cmatrix2.eqn}
  C_{ij} &=& \frac{1}{(x_{i}-x_{j})^{2}} \ . 
\end{eqnarray}

The eigenvectors of $C$ are the normal modes of the random matrix spectrum.  
Each of these eigenvectors describes a statistically independent 
mode of correlated motion of the eigenvalues of the random matrix.  Clearly, 
they provide the natural basis in which to describe the behaviour of 
$P_{N \beta}$ in the vicinity of its maximum.  

The solution to (\ref{equlibrium.eqn}) for an $N \times N$ matrix is 
given by the zeros of the Hermite polynomial, $H_{N}$ 
\begin{equation}
\label{hermite.eqn}
  H_{N}( \sqrt{N} x_{i}) = 0 \ .
\end{equation}
\noindent
This result was obtained by Stieltjes (see appendix A.6 in 
ref.\,\cite{mehta}).  It follows from the observation that Hermite's
differential equation,  
\begin{equation}
  H_{N}''(x) - 2 x H_{N}'(x) + 2 N H_{N}(x) = 0 \ ,
\end{equation}
reduces to (\ref{equlibrium.eqn}) at the zeros of $H_{N}$.  

The results for the eigenvalues and eigenvectors of the matrix $C$ are 
readily stated.  We prove these results in sections 3 and 4.  The eigenvalues 
$\lambda_{k}$ of $C$, which 
satisfy 
\begin{equation}
  \sum_{j=1}^{N} C_{ij} \delta y_{j}^{(k)} = \lambda_{k} 
  \delta y_{i}^{(k)} \ ,    
\end{equation}
are remarkably simple.  Namely,   
\begin{equation}
\label{eigenvalues.eqn}
  \lambda_{k} = - kN \ ,
\end{equation}
where $k$ runs from 1 to $N$.  The $i$-th component of the corresponding 
eigenvector is a polynomial of order $k-1$ evaluated at the equilibrium 
value of $x_{i}$. The first four normalised eigenvectors can be 
written as
\begin{eqnarray}
  \label{1eigenvector.eqn}
  \delta y_{i}^{(1)} &=& \frac{1}{N^{1/2}}
  \\ \label{2eigenvector.eqn}
  \delta y_{i}^{(2)} &=& \left( \frac{2}{N-1} \right)^{1/2} x_{i}           \\
  \delta y_{i}^{(3)} &=& \left(\frac{N-1}{N(N-2)}\right)^{1/2}
                         \left( 1 - \frac{2N}{N-1} x_{i}^{2} \right)        \\
  \label{4eigenvector.eqn}
  \delta y_{i}^{(4)} &=& \left(\frac{2(2N-3)^{2}}{(N-1)(N-2)(N-3)} 
                                                              \right)^{1/2}
                         \left( x_{i} - \frac{2N}{2N-3} x_{i}^{3} \right) \ .
\end{eqnarray}

In the large-$N$ limit, the eigenvector 
component $\delta y_{i}^{(k)}$ is given by the value of the Chebyshev 
polynomial of the second kind, $U_{k-1}(x)$, evaluated at the point 
$x=x_{i}/\sqrt{2}$.  One finds that (up to corrections of order $1/N$)
\begin{equation}
\label{eigenvector.eqn}
\delta y_{i}^{(k)} = N^{-1/2} \ U_{k-1}\left( \frac{x_{i}}{\sqrt{2}}\right)\ . 
\end{equation}

\noindent
This result is natural when one notices that the identity  
\begin{equation}
  \sum_{i=1}^{N} \delta y_{i}^{(k)} \delta y_{i}^{(l)} = \delta_{kl}  
\end{equation} 
can be approximated for large $N$ by 
\begin{equation}
\label{orthogonality.eqn}
  \int_{-\sqrt{2}}^{\sqrt{2}} {\rm d}x \, \rho(x) \ \delta y^{(k)}(x) \ 
  \delta y^{(l)}(x) =   \delta_{kl} 
\end{equation}
with $\rho (x)$ given by (\ref{leveldens.eqn}).  
Eqn.\,(\ref{orthogonality.eqn}) is seen to be the ortho\-go\-nality 
relation for the Chebyshev polynomials of the second kind.

\section{Eigenvalues and Eigenvectors of $C$ for Finite $N$}

To prove relations (\ref{eigenvalues.eqn}) to (\ref{4eigenvector.eqn}), 
begin by looking at the de\-fi\-nition of $C$ stated in (\ref{cmatrix1.eqn}) 
and (\ref{cmatrix2.eqn}), and let $C$ act on a power of $x_i$. We obtain 
\begin{eqnarray}
  \label{eigenvalueequation.eqn}
  \sum_{j=1}^{N} C_{ij} x_{j}^{k} &=& \sum_{j \neq i} 
  \frac{x_{j}^{k}-x_{i}^{k}}{(x_{j}-x_{i})^{2}} - N x_{i}^{k} \\
  &=& \sum_{l=0}^{k-2} (l+1)  x_{i}^{l} \sum_{j \neq i} x_{j}^{k-l-2}
  + kx_{i}^{k-1} \sum_{j \neq i}\frac{1}{x_{j}-x_{i}}- N x_{i}^{k} \nonumber \\
  &=& \sum_{l=0}^{k-2} (l+1) \sigma_{k-l-2} x_{i}^{l} - 
  \frac{1}{2} k(k-1) x_{i}^{k-2} - (k+1) N x_{i}^{k} \nonumber \ .
\end{eqnarray}
Here it is understood that the sums over $l$ are zero for $k \leq 1$ and 
that 
\begin{equation}
  \sigma_{m} \equiv \sum_{j=1}^{N} x_{j}^{m} \ .
\end{equation}
The second identity in (\ref{eigenvalueequation.eqn}) can be proved by
induction, and eqn.\,(\ref{equlibrium.eqn}) can be used to obtain the 
third identity.  It follows from (\ref{eigenvalueequation.eqn}) that the 
$i$-th component of the $k$-th eigenvector, $\delta y_{i}^{(k)}$, is a 
polynomial 
of order $k-1$ in $x_{i}$ and that the corresponding eigenvalue is the 
coefficient, $-kN$, of the highest order term, $x_{i}^{k-1}$.

The sums $\sigma_{m}$ can be determined using the relation
\begin{equation}
  \sum_{i=1}^{N} \sum_{j=1}^{N} C_{ij} x_{j}^{k} 
  = - N \sum_{i=1}^{N} x_{i}^{k} = - N \sigma_{k} \ ,   
\end{equation}
which, together with (\ref{eigenvalueequation.eqn}), lead to the 
recursion relation 
\begin{equation}
\label{recursion.eqn}
   N k \sigma_{k} = \sum_{l=0}^{k-2} (l+1) \sigma_{k-l-2} \sigma_{l} -
  \frac{1}{2} k(k-1) \sigma_{k-2} \ . 
\end{equation}
One observes directly that $\sigma_{0}=N$.  Since the $x_{i}$ come in pairs 
$\pm x$ (or are $0$), it follows that $\sigma_{2k-1}=0$ for all $k \geq 1$. 
This fact and the recursion relation above permit the determination of 
all of the sums $\sigma_{k}$.

Knowing the values of the sums, $\sigma_{k}$, it is now possible to obtain 
all eigenvectors of $C$ using (\ref{eigenvalueequation.eqn}) and to 
verify directly that the vectors (\ref{1eigenvector.eqn}) to 
(\ref{4eigenvector.eqn}) are indeed eigenvectors of $C$ with the 
eigenvalues stated in (\ref{eigenvalues.eqn}).

\section{The Large-$N$ Limit}

In this section we show that, in the large-$N$ limit, the eigenvectors of $C$ 
are related to the Chebyshev polynomials of the second kind as indicated in 
equation (\ref{eigenvector.eqn}).  Define a generating function $G$ 
by the equation
\begin{equation}
\label{generating.eqn}
  G(x) \equiv \sum_{k=0}^{\infty} \sigma_{2k} x^{2k} \ , 
\end{equation}
where the $\sigma_{k}$ are now defined by the large-$N$ form of equation 
(\ref{recursion.eqn}) which reads
\begin{equation}
  N k \sigma_{k} = \sum_{l=0}^{k-2} (l+1) \sigma_{k-l-2} \sigma_{l} \ . 
\end{equation}
Multiplication by $x^{k-1}$ and subsequent summation over $k$ leads to an 
equation which is equivalent to the following differential equation for $G$
     
\begin{equation}
  N \frac{{\rm d}}{{\rm d}x} G = (x G) \frac{{\rm d}}{{\rm d}x} (x G) \ .  
\end{equation}
Since $\sigma_{0}=N$, it follows that $G(0) = N$, and that 

\begin{equation}
  G(x) = N \ \frac{1-\sqrt{1-2x^{2}}}{x^{2}} \ .
\end{equation}
From the definition of $G$ in expression (\ref{generating.eqn}), one observes 
that the $\sigma_{2k}$ are simply
\begin{equation}
  \sigma_{2k} = N \ \frac{(2k-1) \ !!}{(k+1) \ !} \ .
\end{equation}
These values of $\sigma_{2k}$ can also be obtained from the integral
\begin{equation}
\label{sums.eqn}
  \sigma_{2k} = \int_{-\sqrt{2}}^{\sqrt{2}} {\rm d}x \ x^{2k} \rho(x) \ ,
\end{equation}
where $\rho(x)$ is the Wigner semi-circle describing the average level 
density (\ref{leveldens.eqn}).

With these expressions for the sums $\sigma_{2k}$, one can construct the 
eigenvectors of $C$ using a Gram-Schmidt orthogonalisation 
procedure starting with a non-orthogonal basis of vectors with elements 
$x_{i}^{k}$. The coefficients to be determined in the Gram-Schmidt 
procedure are  
\begin{equation}
a_{kj} = - \sum_{i=1}^{N} x_{i}^{k-1} \delta y_{i}^{(j)}  
  = - \int_{-\sqrt{2}}^{\sqrt{2}}{\rm d}x \ x^{k-1}\delta y^{(j)}(x)\rho(x) \ .
\end{equation}
The structure of the integrals implies that the $\delta y^{(k)}(x)$ are 
simply the polynomials orthogonal on the interval $[-\sqrt{2}, \sqrt{2}]$ 
with weight $\rho(x)$.  These polynomials are recognised as the Chebyshev 
polynomials of the second kind, appropriately scaled.  This shows that the 
elements of the eigenvectors $\delta y^{(k)}_{i}$ are given by 
(\ref{eigenvector.eqn}) in the large-$N$ limit.

\section{The Number Variance for the Gaussian Ensembles}

The small amplitude, quadratic approximation to $P_{N \beta}$ in terms of its 
normal modes is useful in calculations of long range spectral fluctuation 
measures.  We illustrate this by determining the asymptotic form of the 
number variance for the Gaussian ensembles. The number variance, 
$\Sigma^{2}(L)$, is defined as the variance of the number of eigenvalues 
in an interval of length $L$.  (Here, it is assumed that the spectrum 
has been ``unfolded'' so that the average spacing between adjacent 
eigenvalues is $1$.)  It is well-known that the exact number variance  
$\Sigma_{\beta}^{2}(L)$ for the Gaussian ensembles has the form \cite{review}  
\begin{equation}
\label{sigma2.eqn}
  \Sigma_{\beta}^{2}(L) = \frac{2}{\beta \pi^{2}} \ln L + 
  K_{\beta} + O(L^{-1}) \ ,
\end{equation}
where $K_{\beta}$ is a constant.  The leading logarithmic term in this 
expression is indicative of a rigid sequence of numbers.  This is to be 
contrasted with the linear $L$-dependence which characterises the 
number variance for a sequence of uncorrelated levels.

We wish to reproduce the logarithmic term in the expression for the 
number variance using the normal modes.  Consider an interval of the 
unfolded eigenvalue spectrum from $-L/2$ to $L/2$. For sufficiently 
large $N$, the level density of the original spectrum corresponding to this 
part of the unfolded spectrum has the constant value of 
$\rho(0) = \sqrt{2} N / \pi$.  Within this interval, the equilibrium position 
of the $k$-th eigenvalue is therefore
\begin{equation}
  x_{k}^{(0)} = \frac{\pi k}{\sqrt{2} N} \ .
\end{equation}
Fluctuations in the eigenvalue spectrum will move the $k$-th eigenvalue 
to a new position which can be written as   
\begin{equation}
\label{fluctuations.eqn}
x_{k} = x_{k}^{(0)}+ \sum_{n=1}^{N} \alpha_{n} \delta y^{(n)}_{k}
      = \frac{\pi k}{\sqrt{2} N} + \frac{1}{\sqrt{N}}
\sum_{n=1}^{N} \alpha_{n} U_{n-1}\left(\frac{\pi k}{2N}\right) \ .
\end{equation}
This means that, when fluctuations are present, the eigenvalues at 
the (unfolded) energies $\pm L/2$ will have eigenvalue numbers
\begin{equation}
  k_{\pm} = \pm \frac{L}{2} - \frac{\sqrt{2N}}{\pi} 
  \sum_{n=1}^{N} \alpha_{n} U_{n-1}\left(\pm \frac{\pi L}{4N} \right) \ .
\end{equation}
Since the number of levels in the interval is now $k_{+}-k_{-}$, it follows 
that the number variance can be approximated by the ensemble
averages 
\begin{eqnarray}
  \Sigma_{\beta}^{2}(L) & \approx & \ \langle (k_{+}-k_{-})^{2} \rangle - 
                    \langle k_{+}-k_{-} \rangle^{2} \nonumber \\
                &=& \frac{8N}{\pi^{2}} \sum_{n=1}^{[\frac{N}{2}]}
                    \langle \alpha_{2n}^{2} \rangle \label{numbervar.eqn}
                    U_{2n-1}^{2}\left( \frac{\pi L}{4N} \right) \\
                &=& \frac{4}{\beta \pi^{2}} \sum_{n=1}^{[\frac{N}{2}]} 
                    \frac{1}{n} U_{2n-1}^{2}\left( \frac{\pi L}{4N} \right) 
                    \nonumber \ .
\end{eqnarray}
Here, we have made use of the fact that terms involving Chebyshev polynomials 
of even order cancel and that the averages over the coefficients are
\begin{equation}
\label{alphaaverage.eqn}
 \langle \alpha_{i} \alpha_{j} \rangle \ = \frac{1}{j\beta N} \ \delta_{ij} \ .
\end{equation}
This last result can be understood by expanding the $\delta x_{i}$ in the 
eigenvectors of $C$,
\begin{equation}
  \delta x_{i} = \sum_{k=1}^{N} \alpha_{k} \delta y_{i}^{(k)} \ .
\end{equation}
The joint probability distribution can now be interpreted as the 
distribution of the $\alpha_{k}$.  In the approximation 
(\ref{approximation.eqn}), this distribution becomes
\begin{eqnarray}
  P_{N\beta}(\alpha_{1}, \alpha_{2}, \ldots, \alpha_{N}) & = & P_{N\beta}^{0} \
  e^{ \frac{1}{2} \beta \sum_{i,j} \delta x_{i} C_{ij} \delta x_{j} } 
\nonumber \\
{} & = & P_{N\beta}^{0} \ e^{ \frac{1}{2} \beta \sum_{k} \lambda_{k} 
\alpha_{k}^{2} }  \nonumber \\
{} & = & P_{N\beta}^{0} \ \prod_k \  
e^{\frac{1}{2}\beta \lambda_k \alpha_{k}^{2}} \ ,
\label{adj5}
\end{eqnarray}
from which it is clear that the $\alpha_{k}$ are independent and Gaussian 
distributed with variance (\ref{alphaaverage.eqn}). 

To calculate the sum over the Chebyshev polynomials, we introduce a 
new variable $\theta$ defined as $\cos \theta = \pi L/4N$ for $L<4N$ and set 
$K \equiv [N/2]$. The terms in the sum can be rewritten as
integrals, and after an interchange of integration and summation we obtain

\begin{eqnarray}
  \sum_{n=1}^{K} \frac{1}{n} U_{2n-1}^{2}\left( \frac{\pi L}{4N} \right) 
  &=&\frac{2}{\sin^{2}\theta} \int_{0}^{\theta} {\rm d} \theta ' \
     \frac{\sin 2K \theta' \sin 2(K+1)\theta'}{\sin 2\theta'} \ .
\end{eqnarray}
The value $L=0$ corresponds to $\theta=\pi/2$.  For this value, the integral 
equals zero.  We can now express the approximate number variance as 
\begin{equation}
 \Sigma_{\beta}^{2}(L) \approx \frac{8}{\beta \pi^{2}\sin^{2}\theta}
 \int_{0}^{\frac{\pi}{2}-\theta} 
 {\rm d} \theta' \ \frac{\sin 2K\theta' \sin 2(K+1)\theta'}{\sin 2\theta'} \ . 
\end{equation} 
For fixed $L$ and large $K$, we see that 
\begin{equation}
  \frac{\pi}{2}-\theta = \frac{\pi L}{4(2K)} \ .
\end{equation}
In this limit, our approximation to the number variance can be written as
\begin{equation}
 \Sigma_{\beta}^{2}(L) \approx 
\frac{4}{\beta \pi^{2}} \int_{0}^{\frac{\pi L}{4}} 
 {\rm d} x \ \frac{\sin^{2} x}{x} = \frac{2}{\beta \pi^{2}}
[\ \ln L- {\rm Ci}\left(\frac{\pi L}{2}\right)+\ln \frac{\pi}{2}+\gamma \ ]\ ,
\label{adj1}
\end{equation}
where ${\rm Ci}\left( \pi L / 2 \right)$ is the cosine integral and 
$\gamma$ is Eulers constant.  For large values of $L$, this function can 
be approximated by 
\begin{equation}
\label{gaussnumber.eqn}
  \Sigma_{\beta}^{2}(L)\approx \frac{2}{\beta \pi^{2}}
  (\ \ln L+\ln\frac{\pi}{2}+\gamma \ ) \ .
\end{equation}
This expression contains a logarithmic term identical to that in 
(\ref{sigma2.eqn}).  The leading term in the number variance for 
large $L$ is thus obtained correctly by the Gaussian approximation to 
$P_{N \beta}$.  The Gaussian approximation is not sufficient to 
reproduce the constant term in (\ref{sigma2.eqn}).

\section{A Sequence of Uncorrelated Levels}

It is useful to find a similar description of the normal modes for a 
sequence of uncorrelated levels.  In this case, it is easiest to proceed 
by considering the correlation matrix 
\begin{equation}
\label{dmatrix.eqn}
  D_{ij} = \langle \ (x_{i} - \langle x_{i} \rangle) \
                   (x_{j} - \langle x_{j} \rangle) \ \rangle
         = \langle x_{i} x_{j} \rangle - 
           \langle x_{i} \rangle \langle x_{j} \rangle \ ,
\end{equation}
where $\langle x_{k} \rangle$ is the ensemble average of eigenvalue $k$. This
approach is slightly different from that adopted above when we considered a
quadratic approximation to $\ln P_{N\beta}$. To the extent that this
approximation is exact, the two approaches lead to identical eigenvectors and
eigenvalues which are negative reciprocals of one another. 

A spectrum of $N$ uncorrelated levels with unit mean level density has a 
Poisson distribution for the level spacings.  The joint probability 
distribution for the levels can thus be written as a product of Poisson 
distributions and step functions:

\begin{eqnarray}
  P_{N}(x_{1}, x_{2}, \ldots, x_{N}) 
  &=& e^{-x_{1}} \ \theta(x_{1}) \  
      \prod_{i=1}^{N-1} e^{-(x_{i+1}-x_{i})} \ \theta(x_{i+1}-x_{i}) 
      \nonumber \\
  &=& e^{-x_N} \ \theta (x_1 ) \ \prod_{i=1}^{N-1} \ \theta(x_{i+1}-x_{i}) \ .
\end{eqnarray}
This form of the joint probability distribution leads immediately to 
the ensemble ave\-ra\-ges $\langle x_{i} \rangle$ and $\langle x_{i} x_{j} 
\rangle$ from which the elements of the matrix $D$ follow:
\begin{equation}
   D_{ij} = \min{\{i,j\}}\ . 
\end{equation}
The eigenvalues, $\omega_{k}$, and eigenvectors, $\psi_{i}^{(k)}$
\hspace{-0.2cm}, of $D$ 
can be obtained by exploiting the special structure of $D$ and expressing the 
eigenvalue problem in the following suggestive manner:
\begin{eqnarray}
 - \ \psi_{1}^{(k)} &=& \omega_{k} \ ( \psi_{2}^{(k)} - 2 \psi_{1}^{(k)})  \\
 - \ \psi_{i}^{(k)} &=& \omega_{k} \ ( \psi_{i-1}^{(k)} - 2 \psi_{i}^{(k)} + 
                                       \psi_{i+1}^{(k)} )                  \\  
     \psi_{N}^{(k)} &=& \omega_{k} \ ( \psi_{N}^{(k)} - \psi_{N-1}^{(k)} ) \ , 
\end{eqnarray}
where $2 \leq i \leq N-1$.  Introducing the definition $\phi_{k} \equiv 
(2k-1)\pi/(2N+1)$, it is readily verified that the normalised eigenvectors 
can be written as 
\begin{equation}
  \psi_{j}^{(k)} = \frac{2}{\sqrt{2N+1}} \sin (j\phi_{k}) \ .  
\end{equation}
The corresponding eigenvalues are
\begin{equation}
\label{poissoneigen.eqn}
  \omega_{k} = \frac{1}{4} \sin^{-2} \left( \frac{\phi_{k}}{2} \right) \ . 
\end{equation}
To facilitate comparison with the random matrix results obtained above, it is 
useful to multiply the $\omega_k$ by  $[\, \rho(0) \, ]^{-2} = \pi^2/2 N^2$ to 
establish identical scales.  We then see that the hardest normal mode 
eigenvalue for the uncorrelated levels is of order 
$-2N^{2}/(\pi^{2}\omega_{N})= -(8 / \pi)^2 N^2$ 
which is comparable to the hardest eigenvalue of $\lambda_{N} = -N^2$ 
obtained for 
the random matrix ensembles.  (The difference is not significant and is 
largely due to the limitations of the quadratic approximation for the joint 
probability density.)  Significant differences are found in the nature of 
the soft spectrum.  For small values of $k$, the uncorrelated levels reveal 
a quadratic spectrum, $-2 k^2$, which is in sharp contrast to the linear 
spectrum, $-kN$, of the random matrix ensembles seen in expression 
(\ref{eigenvalues.eqn}).  

The number variance for a sequence of uncorrelated levels is well known 
\cite{review}
\begin{equation}
\label{poissonnumber_exact.eqn}
  \Sigma^{2}(L) = L \ .
\end{equation}
Here, we wish to proceed in the spirit of the previous section and reproduce 
this result using the normal models just obtained.  Considering the interval 
$[0, L]$ and repeating the arguments which led to equation 
(\ref{numbervar.eqn}), we arrive at the expression
\begin{equation}
\label{poissonnumber.eqn}
  \Sigma^{2}(L) \approx \frac{1}{2N+1}
  \sum_{k=1}^{N} \frac{1}{\sin^{2}(\phi_{k}/2)} \sin^{2}(L\phi_{k}) \ .
\end{equation}
This sum can be performed exactly when $L$ is an integer or a half-integer.  
For these cases, we find that the number variance is given as 
\begin{equation}
  \Sigma^{2}(L) 
  \approx \left\{ \begin{array}{ll} 
                  L  & \textrm{for integer} \ L \\
                  L - \frac{1}{2(2N+1)} & \textrm{for half-integer} \ L 
                  \end{array} 
         \right.
\end{equation}
for physically interesting values $L < N$.  This result agrees with 
(\ref{poissonnumber_exact.eqn}) in the large-$N$ limit.  For other values of 
$L$, it is useful to approximate the sum in (\ref{poissonnumber.eqn}) by an 
integral as in eqn.\,(\ref{adj1}).  
\begin{equation}
\Sigma^{2}(L) \approx \frac{1}{\pi} \, \int_{0}^{\pi/2} \, {\rm d} x \, 
\frac{\sin^2 {(2 L x)}}{x^2} = \frac{1}{\pi^2} \left[ \, 2 \pi L \,
{\rm Si}(2 \pi L) -1 + \cos{(2 \pi L)} \, \right] \ , 
\label{adj2}
\end{equation}
where ${\rm Si}(2 \pi L)$ is the sine integral.  For large values of 
$L$, this leads to the approximation
\begin{equation}
\Sigma^{2}(L) \approx L - \frac{1}{2 \pi^3 L} \sin{(2 \pi L)} - 
\frac{1}{\pi^2} \ ,  
\label{adj3}
\end{equation}
where we note that the final term of $-1/\pi^2$ is an artifact of having 
replaced $\sin^2{x}$ by $x$ in the denominator of (\ref{adj2}).  Evidently, 
eqns.\,(\ref{adj2}) and (\ref{adj3}) invite comparison with the results 
of eqns.\,(\ref{adj1}) and (\ref{gaussnumber.eqn}) obtained for the Gaussian 
ensembles.  We see that the quadratic approximation to the joint 
probability density also provides a reliable description of the number 
variance in the large $L$ limit for uncorrelated levels. 

\section{Discussion and Conclusions}

We have considered the small amplitude normal modes describing the 
fluctuations of the eigenvalues of random matrices about their equilibrium 
positions.  In the limit of large matrices, these modes are essentially plane 
waves.  (Recall that the Chebyshev polynomials have the form 
$U_{2n}(x) = \cos{(2n+1)x}$ and $U_{2n+1}(x) = \sin{(2n+2)x}$ in the limit of 
large $n$ and fixed $x$.)  The mean square amplitude for each mode is 
inversely proportional to its associated eigenvalue, see
eqn.\,(\ref{adj5}).  
Since these eigenvalues grow monotonically with wave number for both the 
Gaussian ensembles and for uncorrelated levels, longer wave length 
fluctuations have a larger amplitude.  Thus, the most probable fluctuation in 
a random matrix spectrum corresponds to a common shift of all eigenvalues 
with no change in their relative separation, see (\ref{1eigenvector.eqn}).
The next most probable fluctuation is a simple ``breathing mode'' of 
the spectrum, see (\ref{2eigenvector.eqn}).  

The properties of the normal mode spectrum provide us with some insight 
regarding the qualitative behaviour of long-range spectral measures.  
For uncorrelated levels, the long wave length spectrum is particularly soft 
with a quadratic spectrum.  As a result, $\Sigma^2 (L)$ is completely 
dominated by soft modes when $L$ is large.  This is seen most easily from 
eqn.\,(\ref{adj2}).  The linear asymptotic behaviour of the number 
variance for uncorrelated levels is thus seen to be a direct consequence 
of the quadratic dispersion relation obeyed by the soft modes.  The situation 
is qualitatively different in the case of the Gaussian ensembles where all 
modes obey an exact linear dispersion relation.  From eqn.\,(\ref{adj1}) 
we see that this linearity ensures that $\Sigma^2 (L)$ must grow 
logarithmically for large $L$ and that all normal modes contribute 
democratically to this asymptotic behaviour. Similar linear dispersion
relations characterise perfectly elastic solids, and it seems useful to
regard the spectral rigidity of random matrices as a consequence of the
physical rigidity of a classical one-dimensional array of line charges.
Dyson was led to the same conclusion \cite{dyson}. Although aware of the
somewhat arbitrary nature of distinctions between phases in one dimension, he
felt it appropriate to call the Coulomb gas (and hence the spectrum of a
random matrix) a ``crystal''. The considerations presented here provide
additional support for this designation. In the same spirit, the quadratic
dispersion relation for the normal modes of uncorrelated levels leads us to
regard them as a gas.


We have emphasised that fluctuations in random matrix spectra are most 
naturally described as the highly correlated motion of individual 
eigenvalues, i.e., the normal modes.  Given the product form of 
eqn.\,(\ref{adj5}), it is also clear that these normal modes are 
statistically independent.  If the matrix $D$ of eqn.\,(\ref{dmatrix.eqn}) 
is calculated in numerical simulations, the statistical errors associated 
with its eigenvalues will become uncorrelated as the sample size increases.  
Soft modes (with large amplitudes) and their eigenvalues can thus be 
determined accurately with relative ease.  This fact can be useful in 
numerical simulations of ensembles which do not readily permit analytic 
analysis.

\section*{Acknowledgements}

We appreciate useful comments on the manuscript by J. Christiansen and
K. Splittorff.


\begin{thebibliography}{10}

\bibitem{review}
T. Guhr, A. M\"uller-Groeling, and H. A. Weidenm\"uller: Random-Matrix
Theories in Quantum Physics: Common Concepts, Physics Reports {\bf 299}, 189
(1998). 

\bibitem{dyson}
F. J. Dyson, Statistical Theory of the Energy Levels of Complex Systems, I,
\linebreak 
J.\,Math.\,Phys.\ {\bf 3}, 140 (1962); Statistical Theory of the 
Energy Levels of Complex Systems. II, J.\,Math.\,Phys. {\bf 3}, 157 (1962).

\bibitem{mehta}
M. L. Mehta, Random Matrices, 2nd Edition, Academic Press (1991). 

\end{thebibliography}
\end{document}